\documentclass{aa}  

\usepackage{graphicx}

\usepackage{txfonts}
\usepackage{hyperref}
\usepackage{xcolor}

\begin{document}

   \title{\textit{NICER} observations reveal doubled timescales in Ansky's quasi-periodic eruptions (QPEs)}

%   \subtitle{I. Overviewing the $\kappa$-mechanism}

\titlerunning{Ansky QPEs 2025}
\authorrunning{Hernández-García et al.}

   \author{L. Hernández-García\fnmsep\thanks{lorena.hernandez@uv.cl}
          \inst{1,2,3,4},
          P. Sánchez-Sáez\inst{5}
          \and
          J. Chakraborty\inst{6}
          \and
          J. Cuadra\inst{7,1}
           \and
          G. Miniutti\inst{8}
          \and
          R. Arcodia\inst{6}\thanks{NHFP Einstein Fellow}
          \and
          P. Ar\'evalo\inst{9,1,2}
          \and
          M. Giustini\inst{8}
          \and
          E. Kara\inst{6}
          \and
          C. Ricci\inst{10,3,11}
          \and
          D. R. Pasham\inst{6}
          \and
          Z. Arzoumanian\inst{12}
          \and
          K. Gendreau\inst{12}
          \and
          P. Lira\inst{13,1}
          }

   \institute{Millennium Nucleus on Transversal Research and Technology to Explore Supermassive Black Holes (TITANS), Gran Breta\~na 1111, Playa Ancha, Valpara\'iso, Chile\\
              \email{lorena.hernandez@uv.cl}
         \and
         Millennium Institute of Astrophysics (MAS), Nuncio Monseñor Sótero Sanz 100, Providencia, Santiago, Chile
         \and
         Instituto de Estudios Astrof\'isicos, Facultad de Ingenier\'ia y Ciencias, Universidad Diego Portales, Av. Ej\'ercitoLibertador 441, Santiago, Chile
         \and
         Centro Interdisciplinario de Data Science, Facultad de Ingenier\'ia y Ciencias, Universidad Diego Portales, Av. Ej\'ercito Libertador 441, Santiago, Chile
         \and
         European Southern Observatory, Karl-Schwarzschild-Strasse 2, 85748 Garching bei München, Germany
         \and
         MIT Kavli Institute for Astrophysics and Space Research, Massachusetts Institute of Technology, Cambridge, MA 02139, USA
         \and   Departamento de Ciencias, Facultad de Artes Liberales,Universidad Adolfo Ibáñez, Av.\ Padre Hurtado 750, Viña del Mar, Chile
         \and
         Centro de Astrobiología (CAB), CSIC-INTA, Camino Bajo del Castillo s/n, 28692 Villanueva de la Cañada, Madrid, Spain
         \and
        Instituto de F\'isica y Astronom\'ia, Facultad de Ciencias, Universidad de Valpara\'iso, Gran Breta\~na 1111, Playa Ancha, Valpara\'iso, Chile
         \and
         Department of Astronomy,University of Geneva, ch. d’Ecogia 16, 1290, Versoix, Switzerland 
         \and
         Kavli Institute for Astronomy and Astrophysics, Peking University, Beijing 100871, People’s Republic of China
         \and
         X-ray Astrophysics Laboratory, Code 662, NASA Goddard Space Flight Center, Greenbelt, MD 20771, USA
         \and
         Departamento de Astronomía, Universidad de Chile, Casilla 36D, Santiago, Chile
             }

   \date{Received September 15, 1996; accepted March 16, 1997}

% \abstract{}{}{}{}{} 
% 5 {} token are mandatory
 
  \abstract
{Quasi-periodic eruptions (QPEs) are recurring X-ray bursts originating from the vicinity of supermassive black holes, but their driving mechanisms 
remain under debate. This study analyzes new \textit{NICER} observations of QPEs in Ansky (a transient event in the nucleus of the galaxy SDSS J1335+0728), taken between January and June 2025. By examining flare durations, peak-to-peak recurrence times, and profiles, we compare the 2025 data with those from 2024 to investigate changes in energy, timescales, and flare shapes. The 2025 QPEs are found to be four times more energetic, with recurrence times of approximately 10 days and flare durations ranging from 2.5 to 4 days, making them both about twice as long as in 2024. Additionally, the flare profiles have become more asymmetric, showing longer decays. 
We explore different theoretical scenarios to explain the observed properties of the QPEs in Ansky, including evolving stream–disk interactions in an extreme mass-ratio inspiral (EMRI) system as a potential mechanism behind the observed changes in recurrence time and energetics, while also considering alternative models based on mass transfer and accretion disk instabilities. Continued observational efforts will be crucial for unveiling the nature of Ansky.
}


   \maketitle
%
%-------------------------------------------------------------------

\section{Introduction}

Quasi-Periodic Eruptions (QPEs) are luminous ($L\sim 10^{42-44}$\,erg\,s$^{-1}$) recurring ($P\sim2.5 h-4.5\,$days) fast ($\Delta T \sim 0.5 h -1.5\,$days) and soft X-ray transients from supermassive black holes (SMBHs) in galactic nuclei \citep{Miniutti19,Giustini20,Arcodia21,Chakraborty21,Quintin23,Arcodia24b,Nicholl24,Chakraborty25a,Hernandez25,Arcodia25}. Despite dedicated searches for QPE counterparts in radio and UV bands \citep{Miniutti19,Giustini24,Wevers25}, 
QPEs have so far been uniquely observed in soft X-rays. They exhibit thermal X-ray spectra with evolving temperatures of $kT\sim 80-250$ eV, and they are typically superimposed on quiescence emission of a disk blackbody with temperature $T_{\mathrm{in}} \sim 40-80$ eV. All QPEs exhibit a characteristic counter-clockwise hysteresis cycle in the $L_{\rm bol}-T$ plane with rises hotter than decays. Assuming a simple blackbody model for the QPE emission, the evolution suggests an expanding emitting region with typical size of the order of one solar radius \citep{Arcodia22,Miniutti23a,Quintin23,Giustini24,Nicholl24,Hernandez25,Chakraborty25a}. The timing behavior of the different QPE sources is diverse, with some sources showing a regular pattern (e.g. GSN~069 or eRO-QPE2, see \citealt{Miniutti23a} and \citealt{Arcodia24c}) and others having more complex or even erratic time series (e.g. RX~J1301.9+2747 or AT2022upj, as discussed by \citealt{Giustini24} and \citealt{Chakraborty25a}).

Several lines of evidence point towards a connection between QPEs and transient events such as Tidal Disruption Events (TDEs), suggesting that QPEs are detected  after the TDE optical/UV peak, possibly in as many as $9^{+9}_{-5}$\% of TDEs \citep{Chakraborty25a}, although the latter estimate is subject to large uncertainties mostly due to the still low number statistics. The detection of QPEs months to years after a transient event, AT2019vcb, XMMSL1~J024, AT2019qiz, AT2022upj, and ZTF19acnskyy (`Ansky') is among the  clearest evidence for a QPE-transient event connection \citep{Miniutti23a,Guolo25,Quintin23,Bykov25,Chakraborty21,Nicholl24,Chakraborty25a,Hernandez25}.

The QPE physical origin is still uncertain and subject of active research. Models proposed so far cluster in two main categories, namely those that invoke different flavors of disk instabilities \citep{Raj21,Pan22,Sniegowska23,Kaur23,Pan23} and those associated with the interaction between the central SMBH (and/or its disk) and a secondary orbiting companion within an extreme mass-ratio inspiral (EMRI) system. The latter class  further divides into models invoking mass transfer at pericenter from the companion star \citep{King20,King22,Zhao22,Wang22,Krolik22,LinialSari23,Chen23,Lu23,Olejak25} and those that call on collisions between the companion (a star or stellar mass black hole) and the (likely post-TDE or a new accretion flow) accretion disk, in which QPEs result either from the collision itself \citep{Sukova21,Xian21,Franchini23,Linial23,Tagawa23,Zhou24,Vurm24} or from shocks due to the impact of a stream of debris ablated from the star at each collision with the disk itself.  Which of the latter two effects dominates QPE emission is a function of the disk (and stream) properties at the impact site \citep{YaoP25}. Recent works suggest that debris-disk shocks might indeed represent a preferential channel for QPE emission especially for long-period QPE systems (such as Ansky) where the interaction is likely to occur at large radii and relatively low disk density \citep{YaoP25,mummery2025,linial2025}.

If QPEs are indeed produced by EMRIs interacting with post-TDE disks or experiencing mass transfer episodes, this would have the implication
that QPEs may represent the first-observed electromagnetic counterpart to EMRIs, potentially detectable as multi-messenger sources by future millihertz gravitational wave detectors such as \textit{LISA} \citep{Amaro-Seoane2023} and \textit{TianQin} \citep{Luo16}.

\cite{Hernandez25} recently reported the discovery of QPEs in the nucleus of the galaxy SDSS J1335+0728, at a redshift of 0.024, and with an estimated black hole mass of $\sim 10^6 M_\odot$, based on its stellar mass \citep{Reines15}. This followed an optical transient, named Ansky after its ZTF object ID (ZTF19acnskyy), which is consistent with either a `turning on' active galactic nucleus (AGN) or an unusual nuclear transient \citep{Sanchez24}.
The observed QPEs exhibited a peak luminosity of $L\sim 2\times 10^{43}$\,erg\,s$^{-1}$, had typical duration of $\sim$1.5 days, and were separated by $\sim$4.5 days. 
At $F_{\mathrm{peak}}>10^{-11}$\,erg\,s$^{-1}$\,cm$^{-2}$, it is 10 times brighter than any known QPE.
One remarkable detail of this event was the initial lack of X-ray emission, constrained by eROSITA (2020--2023) and \textit{Swift}/XRT upper limits (Aug. 2021 -- Jul. 2023).
In 2024, Ansky showed the longest durations and recurrence times of all known QPEs, with the most extreme timing behavior seen yet. 
Ansky also showed significant X-ray spectral, time-varying emission/absorption feature changing in centroid, width, and depth over the course of each eruption, which was modeled by \cite{Chakraborty25b} as a time-evolving P Cygni profile, possibly emerging from the debris ejected by the orbiter-disk collision \citep{Vurm24}. 

Given its brightness, Ansky is an ideal target for continued monitoring to shed light on the evolution of QPEs with time. Over the past year, \textit{NICER} has observed the source extensively, providing valuable data for tracking these changes.
In this work, we report recent (2025) \textit{NICER} follow-up observations of Ansky, which revealed that it has changed its eruption pattern since the last observations obtained in July 2024. Section \ref{data} shows the data used in this work. Section \ref{results} presents the results of the data analysis. In Section \ref{discussion}, we discuss potential mechanisms to explain the new observations. Finally, in Section \ref{summary}, we summarize our findings.

%--------------------------------------------------------------------

\begin{figure*}
    \centering
    \includegraphics[width=\textwidth]{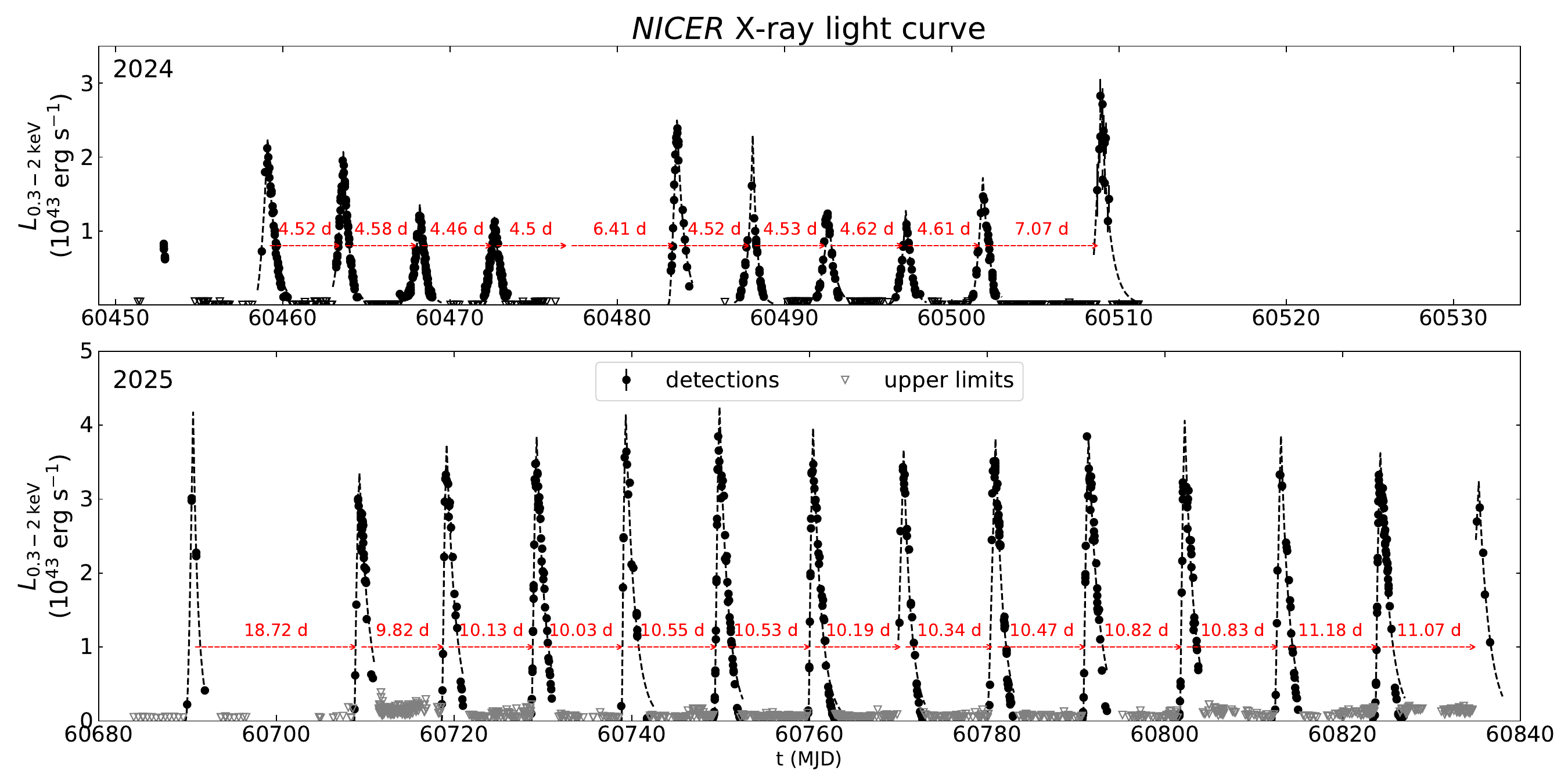}
    \caption{\textit{NICER} X-ray light curve of Ansky for the period (upper panel) May 19--July 20, 2024. Detections are plotted as black circles, and non-detections as gray triangles. The typical peak-to-peak timescale is $\sim$4.5 days; these QPEs were presented in \cite{Hernandez25}. \textit{NICER} light curve (bottom panel) between January 7 and March 30, 2025. 
    The typical peak-to-peak timescale is $\sim$10 days. Note that there are no observations around MJD 60700, so there might be a missing flare.
    The dashed lines represent exponential rise and decay profiles fit to each QPE (see Sect.~\ref{results}). The horizontal axis spans an equal number of days in 2024 and 2025 to facilitate timescale comparison.
    }
    \label{fig:lc}
\end{figure*}

\begin{figure}
    \centering
    \includegraphics[width=\linewidth]{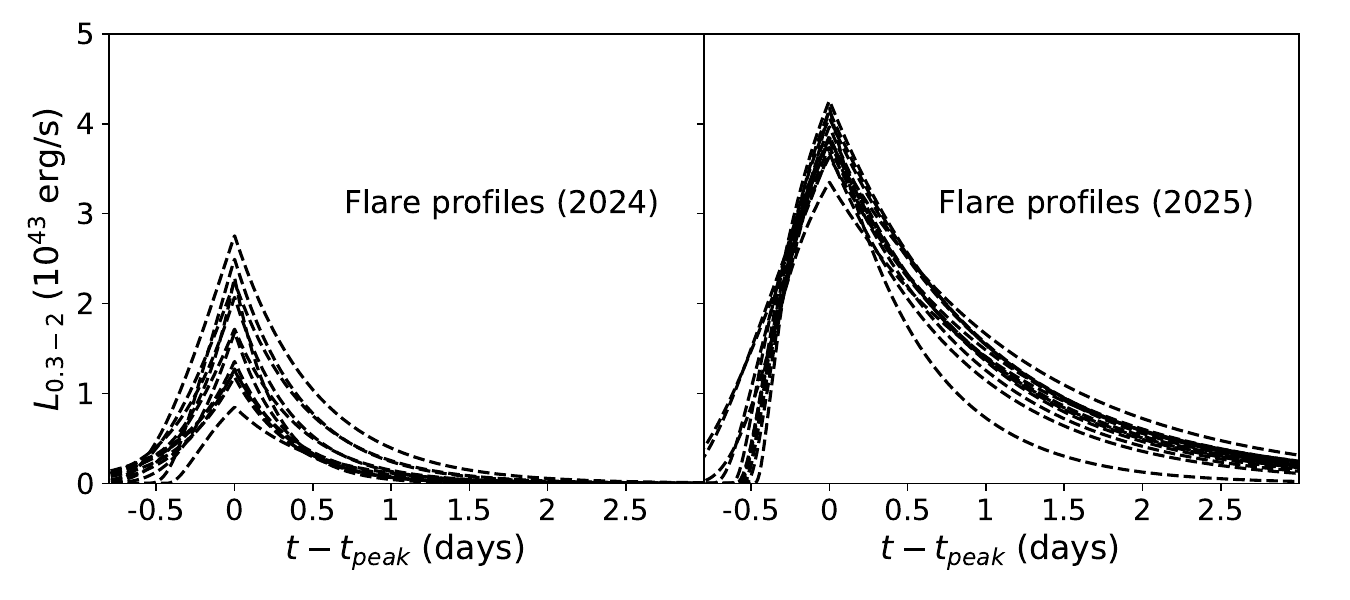}
    \caption{Comparison of the flare profiles from 2024 (left panel) to 2025 (right panel).}
    \label{fig:flare_profiles}
\end{figure}

\begin{figure}
    \centering
    \includegraphics[width=\linewidth]{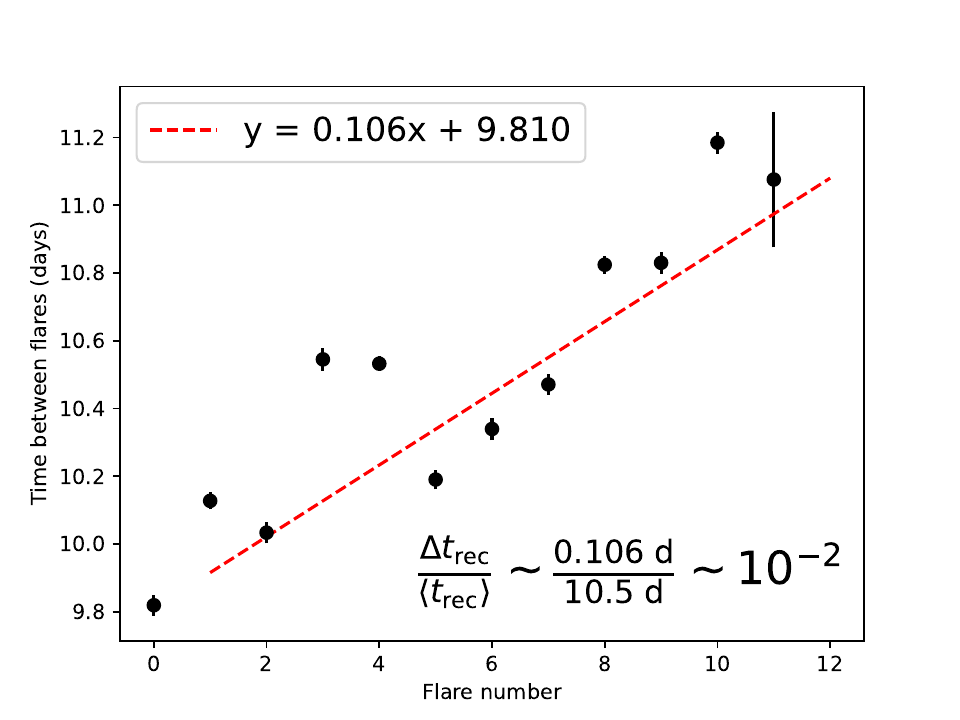}
    \caption{Time between consecutive flares as a function of flare number. Error bars reflect uncertainties propagated from the flare peak times. A linear fit (red dashed line) shows a gradual increase in recurrence time, rising by about 0.1 days per flare.}
    \label{fig:linealfit}%
\end{figure}

\section{Observations and data reduction}\label{data}

The \textit{NICER} X-ray Timing Instrument \citep{Gendreau16} observed Ansky for a total of 283~ks across 213 observations from January 7th to June 16th, 2025, in addition to the 278~ks across 60 observations  from May 19-July 20, 2024 presented in \cite{Hernandez25}. 
No observations were carried out between these dates because the source was behind the Sun. The data were processed using \texttt{HEAsoft} v6.33 and \texttt{NICERDAS} v12. We followed the time-resolved spectroscopy approach for reliably estimating the background-subtracted light curve outlined in \cite{Chakraborty24}; we refer the reader to their Section 2.1 for details. The light curve thus generated is presented in Fig.~\ref{fig:lc}. We consider a source ``detection'' as any Good Time Interval (GTI) in which the addition of a source component is an improvement over the background-only fit by $\Delta$C-stat$\geq 25$, which was determined empirically as a suitable boundary. All GTIs are 200 seconds. The flux upper limits are in the range $(2.3-4.1)\times$10$^{-13}$\,erg\,s$^{-1}$\,cm$^{-2}$. 
The resulting time-resolved blackbody fit parameters were used to build the plots in Section~\ref{results}.

\begin{figure}
    \centering
    \includegraphics[width=\linewidth]{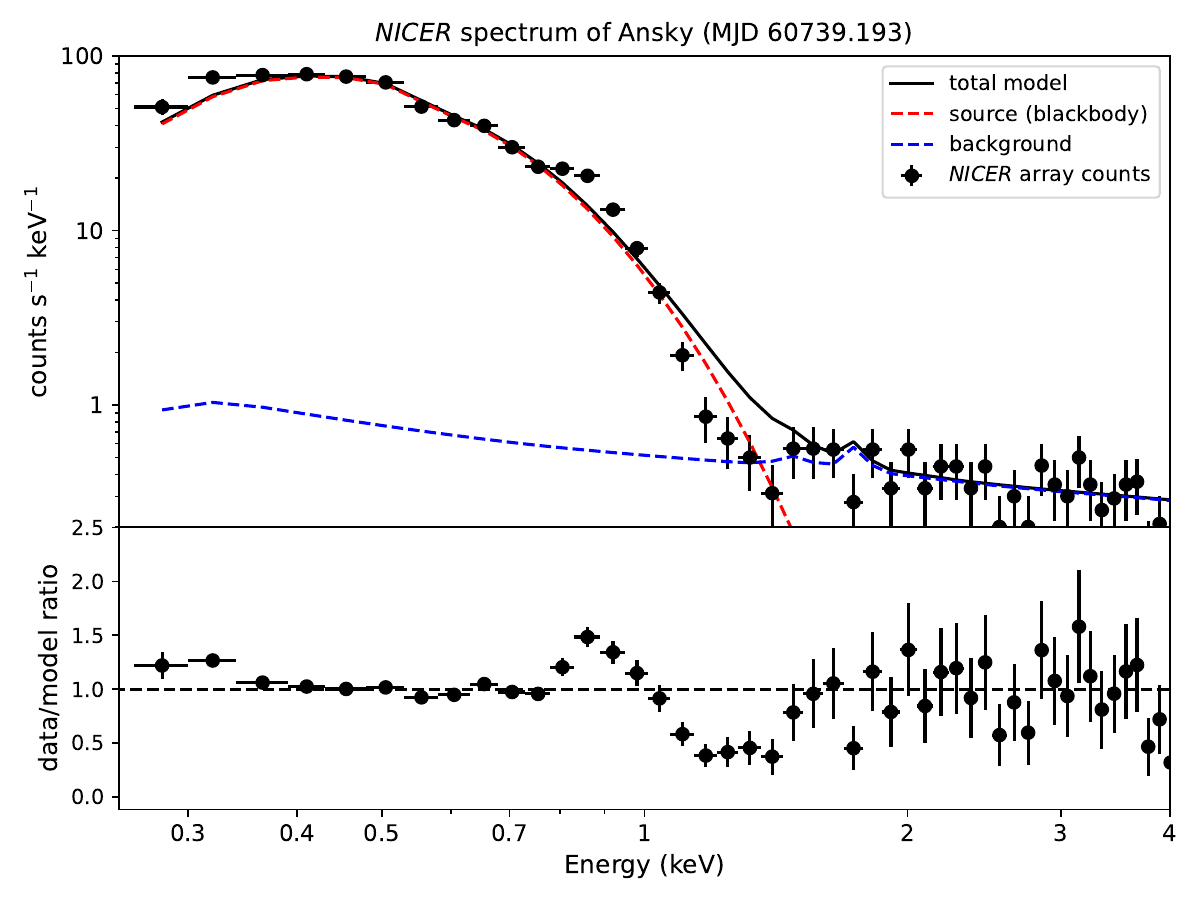}
    \caption{Background-subtracted X-ray spectrum from \textit{NICER}. This is an example of a high-flux spectrum, taken from one 200-second exposure, used to estimate the spectral parameters. The spectrum is fitted with a blackbody model. The residuals in the bottom panel show the presence of an additional component, corresponding to the emission/absorption feature around $\sim$1 keV, analyzed in detail in \cite{Chakraborty25b}.}
     \label{fig:nicer_spec}
\end{figure}
   
\begin{figure*}
    \centering
    \includegraphics[width=\textwidth]{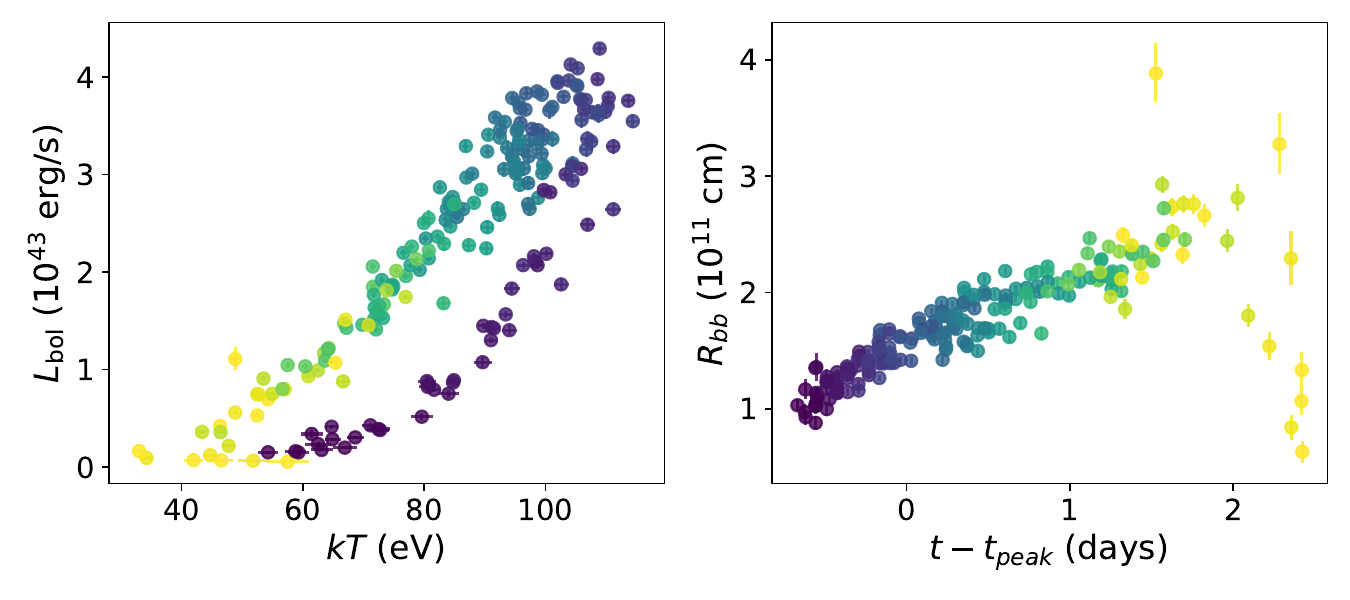}
    \caption{The spectral evolution of the QPEs in Ansky for the bursts observed by \textit{NICER} in 2025. The color scheme indicates the time relative to peak (dark/light at early/late times). The QPEs undergo hysteresis in the $L$-$kT$ plane (left panel), which, for a blackbody-like spectrum, can be interpreted as an expanding emission region $R_{\mathrm{bb}}$ (right panel). The error bars represent $\pm 1\sigma$ uncertainties.}
    \label{fig:hysteresis}
\end{figure*}

\begin{figure}
    \centering
    \includegraphics[width=\linewidth]{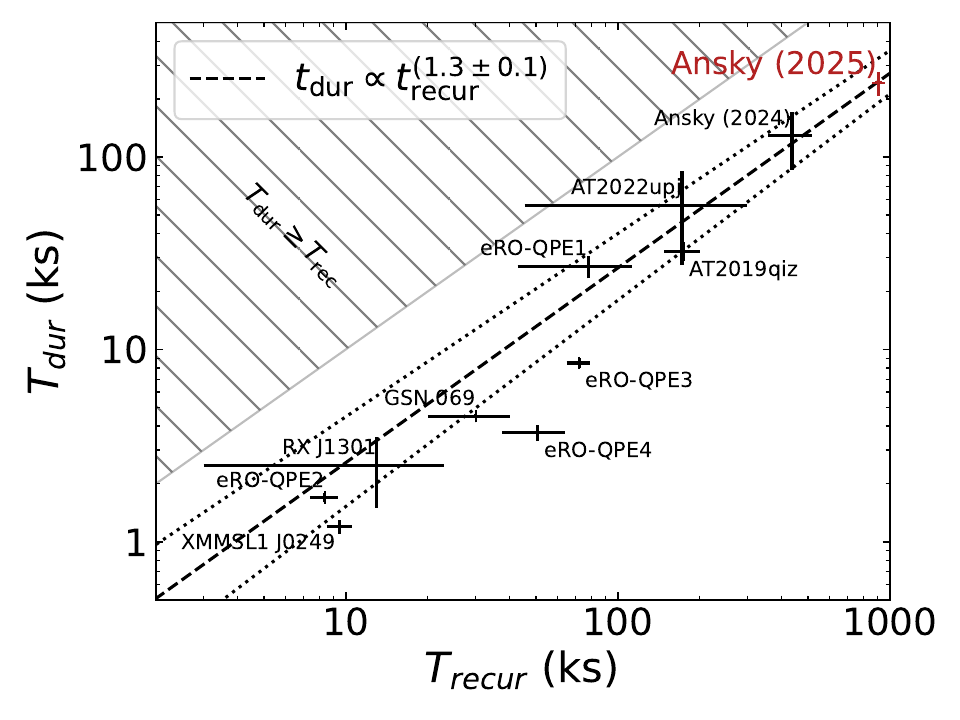}
    \caption{QPE population-level $t_{\rm dur}-t_{\rm recur}$ relation.
    Burst duration and recurrence times are shown for known QPEs: Ansky shows an increase in both timescales compared to QPEs in 2024, but closely follows the power-law fit to the observed data points (dashed line; dotted lines represent one standard deviation). The error bars indicate the range of timescales observed for each source, rather than representing measurement uncertainties.    }
    \label{fig:timescales}%
\end{figure}

\section{Results}\label{results}

In Figure~\ref{fig:lc} (bottom panel), we present the \textit{NICER} X-ray data light curve generated by our time-resolved spectroscopy approach outlined in Section~\ref{data}. In total, 14 QPEs were observed between January and June, with recurrence times (peak-to-peak separations) of $\sim 10$ days, and durations (time spent about $\geq 0.01L_{\rm peak}$) of $\sim 2.1-3.6 $ days. Note that there is a gap in observations around MJD 60700, where a flare was possibly missed.
These timescales are approximately twice as long as those observed during 2024 and reported in \cite{Hernandez25}, when
flare duration was $\sim$1.5 days and peak-to-peak duration was $\sim$4.5 days (Figure~\ref{fig:lc}, top panel).

We used an exponential rise/decay model, which is common in the literature for QPEs \citep{Arcodia22}, to fit the flare profiles. The model has the form:
\[ \mathrm{ F_{\rm QPE} } = \begin{cases} 
      A\lambda e^{{\tau_1/(t_{peak}-t_{as}-t)}} & \mathrm{if}\; t<t_{peak} \\
      Ae^{{-(t-t_{peak})/\tau_2}} & \mathrm{if}\; t\geq t_{peak}
   \end{cases}
\]
where $F$ is the flux of the QPE, $A$ is the flare amplitude; $t_{peak}$ is the time of peak flux; $\tau_1$, $\tau_2$ are the e-folding times of rise and decay, respectively; $\lambda=e^{\sqrt{\tau_1/\tau_2}}$ is a normalization to join rise and decay; and $t_{as}=\sqrt{\tau_1 \tau_2}$ sets the asymptote time such that $F_{\rm QPE}=0$ for $t<t_{peak}-t_{as}$. 

By integrating over the above definition of burst duration, we computed the total energy outputs of the flares, resulting in an average integrated energy output of $(3.98\pm0.4)\times 10^{48}$ erg. This represents a factor four increase with respect to the flares observed in 2024 \citep{Hernandez25}, and at least a couple of orders of magnitude higher than the typical QPE energy in other sources \citep[e.g.,][]{Miniutti23a}.

Flare profiles are presented for comparison in Figure~\ref{fig:flare_profiles}, for the eruptions in 2024 (already presented in \citealt{Chakraborty25b}; left panel) and 2025 (right panel). The flares from 2025 are visibly more energetic than those from 2024, as quantified above. Additionally, the 2025 profiles have a longer duration and exhibit a more asymmetric shape compared to the more symmetric profiles observed in 2024. 
We quantified the flare rise and decay times from the best-fit model of $\mathrm{ F_{\rm QPE} }$, with the rise time being measured as the interval during which the flare brightens from 1/e$^ 3$, where e is the Euler's number, of the peak luminosity to the peak itself, while the decay time corresponds to the interval from the peak back down to 1/e$^ 3$. This threshold was chosen as in \cite{Arcodia22}, but a simpler analysis was performed here.

For 2024, we obtain a mean rise time of 0.81$\pm$0.08 days and a mean decay time of 1.28$\pm$0.10 days. In 2025, the rise time is 0.84$\pm$0.06 days, while the decay time increases significantly to 2.23$\pm$0.15 days. Uncertainties are estimated as the standard deviation of the mean. In both years, the rise time is consistently shorter than the decay time, and while the change in rise time is not statistically significant, the decay time roughly doubles from 2024 to 2025. 

Importantly, this change in flare characteristics is not gradual; while the flares of 2024 show a high degree of similarity to each other, the flares of 2025 also exhibit a high degree of similarity within their group, but when comparing the two years, their shapes differ significantly, indicating a drastic shift in flare behavior in only $\sim$six months. 
We notice that even when commparing the normalized flares, the shapes are different.

When comparing the periodicity of the data, we can estimate  a coarse mean rate of change of period between the 2024 and 2025 observing seasons as $\dot P \equiv \Delta P / \Delta t \approx +0.02$. 
Moreover, at least thirteen consecutive flares are seen at regular intervals as opposed to the behavior in 2024, when longer recurrence times appeared after groups of five flares. 

In Fig. \ref{fig:linealfit} we show the time between consecutive flares as a function of flare number for the 2025 observations, based on the flare peak times estimated from the exponential fits. Each data point corresponds to the interval between two successive flares, and the associated error bars represent the uncertainties propagated from the timing uncertainties of the peak positions. A linear fit to the recurrence times (shown as a red dashed line) reveals a gradual increase in the flare period throughout the observations. The best-fit slope is approximately 0.1 days per flare, suggesting a slow but systematic lengthening of the recurrence time. 
The same analysis cannot be done for the 2024 data, but note that this corresponds to a 
$\dot P \approx +10^{-2}$, i.e., 
of the same order of magnitude as the change observed between the 2024 and the 2025 seasons.
Furthermore, it is worth noting that such an increase in recurrence time has not been commonly observed in other QPE sources, and interpreting this trend, as well as the change in period between 2024 and 2025 is beyond the scope of the present study, since the physical mechanism responsible remains unclear.

Figure~\ref{fig:nicer_spec} shows an example of one of the high signal-to-noise, 200-second exposure, \textit{NICER} spectra. 
The spectral fit of the data was done using the model \texttt{tbabs*zbbody} with frozen Galactic $N_H=2.6\times 10^{20}$ cm$^{-2}$. This results in blackbody temperatures in the range kT = [33-114] eV and bolometric luminosities, which were computed from 0.001--100 keV using the \texttt{clumin} model in \texttt{XSPEC}, in the range L$_{bol}$ = [0.06--4.29]$\times$10$^{43}$\,erg \,s$^{-1}$. The residuals show a clear necessity of an additional component compatible with the P-Cygni profiles reported in \cite{Hernandez25} and \cite{Chakraborty25b}.

In the left panel of Figure~\ref{fig:hysteresis} we present the bolometric luminosities as a function of temperature, illustrating the ``hysteresis'' behavior commonly observed in QPEs, where the rise occurs at higher temperatures than the decline. This spectral evolution is often regarded as a hallmark of QPEs, as it appears in all known sources, suggesting a shared physical mechanism underlying their emission. 
In the right panel of Figure~\ref{fig:hysteresis}, we show the inferred blackbody radius from the Stefan-Boltzmann law, $R_{\rm bb} = \sqrt{L_{\rm bol}/4\pi\sigma_{\rm SB}T^4}$. The
emission radius starts at $R_{bb}\sim R_\odot$ and grows by a factor of 2–-3 over the course of an eruption, as seen in other QPEs \citep{Miniutti23a,Quintin23,Arcodia22,Chakraborty24,Giustini24,Nicholl24}. 
This suggests that the emission region expands significantly as the eruption progresses, with the changes in luminosity and temperature reflecting the evolving size of the emitting region. We note that out-of-equilibrium effects in the photon-starved ejecta could cause the measured blackbody radius to misrepresent the ejecta's physical size \citep{Linial23,Vurm24}. In Ansky we also observe this behaviour, which is qualitatively consistent with other QPEs.

In Fig.~\ref{fig:timescales}, we show the burst durations, $T_{\rm dur}$, with peak-to-peak recurrence time, $T_{\rm rec}$, of Ansky in 2024 and 2025, along with those for other QPEs in the literature. We define the flare duration as the interval beginning when the model flux rises above $0.01L_{\rm peak}$ and ending when it decreases below this threshold. 
With these flare profiles, we find a mean peak-to-peak recurrence time of $t_{\mathrm{recur}} = 907\pm 25$~ks and mean duration of $t_{\mathrm{dur}} = 243 \pm 35$~ks.
We recall that these parameters were $t_{\mathrm{recur}} = 436.8\pm 80.3$~ks and mean duration of $t_{\mathrm{dur}} = 129\pm 43.2$~ks in 2024 \citep{Hernandez25}\footnote{Note that this estimation includes the longer gaps.}.
Ansky fits well in the empirical scaling relation of $T_{\rm dur}\propto T_{\rm rec}^{1.3}$ observed in QPEs \citep{Hernandez25,Chakraborty25a}. It had the longest timescales in 2024, and it can be seen that the new eruptions further extend this relationship. We caution that this relation remains subject to small-number statistics, and gaps in observational coverage may influence this apparent correlation. The shaded regions represent excluded areas of the parameter space. For instance, sources with $T_{\rm dur}\geq T_{\rm rec}$ would not be identifiable as QPEs, as they would lack well-resolved individual bursts,  and note that the observational appearance of flares with $T_{\rm dur}\simeq T_{\rm rec}$ would be that of a quasi periodic oscillation (QPO) rather than a QPE \citep{Dacheng13,Webbe23}.

\section{Discussion}\label{discussion}

The new, 2025 observing campaign reveals QPEs four times more energetic than in 2024, with an average recurrence time of $\sim$ 10 days, and flare durations of 2.5$- $4 days, in both cases about twice as long as observed in 2024. The extreme variation in both timing and radiated energy, occurring in less than one year, represent a challenge for QPE models, some of which are briefly discussed below.

\subsection{EMRI collision models}

In the framework of star-disk collisions in an EMRI system, QPEs might be produced via two different mechanisms, namely the star-disk interaction itself or the interaction between the disk and a debris stream that is ablated from the star at each collision. Assuming for simplicity the case in which the debris remains within the Hill sphere, the dominant mechanism for QPE production depends on the relative disk and debris surface density at the impact site, $\Sigma_{\rm disk}$ and $\Sigma_{\rm debris}$, as discussed in detail by \citet{YaoP25}. For $\Sigma_{\rm debris} \ll \Sigma_{\rm disk}$, the shock between the stream and the disk is inefficient to photon production and direct star-disk collision are expected to dominate. On the other hand, when $\Sigma_{\rm debris} \gtrsim \Sigma_{\rm disk}$ shocks from the debris-disk interaction dominate the QPE emission due to the large size of the stream, its low optical depth, and a prolonged interaction time. If $\Sigma_{\rm debris} \gg \Sigma_{\rm disk}$, the debris stream acts basically as a star and QPEs are produced in a similar fashion as in direct star-disk collision discussed below \citep{YaoP25, linial2025,mummery2025}.

The impact of a star with the disk results in a radiation-mediated shock in which a portion of the shocked disk material is expelled from the mid-plane in a two-sided fountain expanding due to radiation pressure. QPEs are then generated by the photons emerging from the inner photosphere embedded in the expanding ejecta \citep{Linial23,Vurm24,Chakraborty25b}. In general, given the presence of two-sided ejecta with similar properties, one then expects to observe two QPEs per orbit. Moreover, the photons escaping from the inner photosphere need to go through the surrounding expanding ejecta, which can imprint spectral features associated with outflowing material. Such P-Cygni profiles, evolving in strength and velocity during the single QPE, were indeed detected during the 2024 campaign in the QPE X-ray spectra of Ansky, which provides some support for the overall star-disk collision model from 2024, or at least for the QPE emission mechanism in terms of expanding ejecta \citep{Chakraborty25b}. 

Such a model, however, fails to explain the sudden change in timing properties, as well as energetics, from the 2024 to the 2025 campaign. The doubling of recurrence times would imply that the secondary has moved outwards significantly, which seems unfeasible even considering possible interactions with third bodies. Moreover, the star would now impact less dense disk material likely leading to less, rather than more, energetic bursts. A plausible solution would be that the EMRI orbit is still the same but we are now observing only one QPE per orbit instead of two, which is also consistent with the almost precise doubling of recurrence times. This might occur because of geometrical effects in which one collision is missed (a warped, precessing, or elliptical disk, for example). Orbital eccentricity might also result in missed collisions depending on apsidal phase, but the regularity of the recurrence times in both 2024 and 2025 suggests that the orbital eccentricity is nearly zero in Ansky. 

On the other hand, \cite{YaoP25}, \citet{mummery2025}, and \citet{linial2025} suggest that, within the EMRI collision scenario, QPEs might be dominated by impacts between the debris stream ablated from the star at each star-disk collision rather than by the star-disk interaction itself. Debris-disk collision are in fact expected to be more efficient than star-disk ones, especially for sources with long recurrence times ($\gtrsim $ 1 day) and duration, and with large radiated energy, as is the case for Ansky. Moreover, debris-disk interactions account in a more natural way for some of the QPE observed properties (e.g. the correlation between duration and recurrence, see Fig.~6 of \citet{linial2025}, among others). Within this context, and if the stream does not have enough inertia to penetrate the disk, the debris-disk impact is expected to produce only one QPE per orbit  since the emission from the resulting shock is blocked on one side by the optically thick disk. However, streams whose density contrast with the disk is higher might be able to penetrate the disk and give rise to two-sided ejecta similar to those that originate from direct star-disk collision therefore producing two QPEs per orbit \citep{YaoP25}.

A possible qualitative picture therefore emerges: in 2024, the debris penetrate the disk giving rise to two-sided ejecta dominating the X-ray emission, as in the case of star-disk collisions, and producing two QPEs per orbit. This might be, for example, because of a high disk scale-height or low density at the impact site during the early stages post-transient events, or to the coupled evolution of the star-disk system leading to rapid changes in the disk structure \citep[as discussed by][]{Linial24c}. As the disk structure evolves with time, as likely does that of debris streams with collision number, the density contrast between the debris and the disk is reduced in 2025, so that the stream does not penetrate the disk anymore, shock emission is confined to only one side of the disk mid-plane, and only one QPE per orbit is visible, resulting in the observed doubling of recurrence times. A more tenuous debris stream in 2025 is also likely larger in size than in 2024, possibly explaining the longer duration and higher luminosity of QPEs. 
However, it is not immediately clear that a reduction in stream density, even if coupled with a larger extent, can increase the eruption energy by a factor of four as observed, and this is a worthy avenue for further investigation.

The picture outlined above is qualitative only, but could be tested more quantitatively through simulations of star ablation as well as disk structure evolution. We must however point out that the disk in Ansky might not be a pure post-TDE one if an AGN disk was already present, which may introduce further complexity \citep{Sanchez24,Hernandez25}. If the picture we suggest is correct, the recurrence time of Ansky's QPEs should stabilize at $\sim$ 10 days in the near future, although rapid switching between $\sim$ 5 and $\sim$ 10 days cannot be excluded \citep{Linial24c}. The detection of doubling recurrence times in other QPE sources with long recurrence in the future may represent a smoking gun for the proposed scenario.

\subsection{Alternative scenarios: mass transfer and disk instabilities}

QPEs have also been associated with mass transfer scenarios from a white dwarf in a highly eccentric orbit \citep[e.g.][]{King20,King22,Wang22} or from stars (main sequence or evolved ones) on more circular ones \citep[see, for example,][]{Zhao22,Lu23,Olejak25, DOrazio25} with emission being generally attributed to accretion onto the SMBH, perhaps mediated by the existing disk. In all cases, however, mass transfer occurs at pericenter, and pericenter passages are a rather stable clock which is unlikely to experience large variations. The alternating longer and shorter recurrence times that are typical of GSN~069, eRO-QPE2, RX~J1301.9+2747 and other sources are problematic within that context. 
While it has been argued that stable mass transfer could explain these features \citep{King22}, and more recent works extending these models are even able to reproduce the year-long disappearance of flares \citep{Yang25}, the mechanism by which such stability produces a variable recurrence time remains debated \citep[e.g.,][]{LinialSari23}, and the system evolution may likely end with the complete disruption of the star.

Moreover, QPEs are typically observed after a TDE, not before, complicating the viability of pre-existing binary mass transfer as the trigger. For main-sequence donors, \citet{Ryu20iii} suggested that repeated tidal interactions could cause stochastic energy exchange, potentially explaining a positive $\dot{P}$ as seen in Ansky. However, the magnitude of the observed period change would require substantial mass loss, likely altering the stellar structure to an extent incompatible with continued QPE production \citep{Linial24a}.

An alternative hybrid scenario may involve interaction between a pericenter mass stream and a remnant post-TDE disk, resembling a debris–disk or stream–stream collision model \citep[e.g.,][]{Krolik22}. In such a case, variability in the recurrence time could emerge from dynamical changes in the disk rather than the orbit itself. Still, this would require detailed modeling to assess whether the observed properties of Ansky and similar systems can be consistently reproduced.

Accretion disk instability models do not suffer from the strict timing requirements set by an orbiter, and may thus be able to explain changes in recurrence time at different epochs. As the disk evolves, the radial location and extent of the unstable region is bound to change leading to different characteristic timescales. Classical radiation pressure instabilities are associated with timescales that are much too long in SMBH disks to account for observed QPEs. However, the presence of large scale magnetic fields can shrink the unstable region to sufficiently small sizes and radii, for a given range of black hole mass and accretion rate, that the typical hours to days recurrence timescales of QPEs can be reproduced \citep{Pan22,Sniegowska23,Kaur23,Pan23}. Moreover, the spectral evolution during QPEs can also be accounted for as shown, for example, by \citet{Pan23}. 
A recent extension of this framework identifies critical thresholds in accretion rate and magnetic field parameters that separate stable and unstable QPE regimes, naturally explaining both regular and stochastic eruption patterns while maintaining consistent peak temperatures \citep{pan2025}.

 The profile of the single burst is also different from the observed QPEs. While the QPE rise is generally significantly faster than its decay, bursts associated with radiation-pressure instability are characterized by slow rises during the interval between bursts that culminate at peak and are then followed by very fast decays, see e.g., the typical profile of heartbeats in GRS~1915+105 \citep{Neilsen11} or the profiles from simulated light curves of QPEs (Fig.~2, 5, and 6 in \citet{Pan23} for instance), but this is not observed in any of the QPE sources. Other instabilities such as the thermal-viscous instability can produce fast rise and slow decays as seen, for example, in X-ray binary outbursts \citep{Remillard06}, but they are associated with much longer timescales in SMBH disks and do not seem capable of producing recurrence times as short as few hours as observed in some QPEs, even considering magnetically-supported disks \citep{Hameury09,Noda18}, while disc tearing instabilities still need to be applied more quantitatively to QPEs \citep{Raj21}. In summary, the radiation-pressure instability in modified, magnetized disks appears to be a promising alternative mechanism for QPE production that could potentially explain the observed changes in QPE timescales of Ansky, thus should be explored further together with EMRI-based models.

\section{Summary}\label{summary}

In this manuscript we report \textit{NICER} observations of the QPEs in Ansky, taken between January and June 2025. The new observing campaign reveals QPEs four times more energetic than in 2024, with a peak-to-peak recurrence time of $\sim$10 days, and flare durations of 2.5$-$4 days. These time scales are $\approx$two times longer than those observed between May and July 2024. We estimate the mean rate of change of period between both observing seasons as $\dot P  \approx +0.02$. 
Interestingly,  the interval between consecutive flares is increasing at a rate of $\sim$0.1 days per flare during 2025, with a $\dot P  \approx +10^{-2}$.
The flares have also changed their profiles, now showing more asymmetry than in 2024 -- while the rise times remain similar in both years, the decay times in 2025 are roughly twice as long as those observed in 2024.

We discuss potential mechanisms that may explain the new observed timescales.
We propose a qualitative scenario in which the QPEs in Ansky originate from interactions between a stellar companion and a post-TDE accretion disk or a newly formed accretion flow in an EMRI system. 
Specifically, we suggest that the dominant emission mechanism evolved from two-sided shocks caused by debris streams penetrating the disk in 2024 (producing two QPEs per orbit), to one-sided shocks in 2025, as a result of changing disk and stream properties that reduced the stream’s ability to cross the disk. This transition naturally explains the observed doubling of recurrence times, as well as the longer duration and higher luminosity of QPEs in 2025.

Although this picture provides a coherent explanation for the temporal and spectral evolution of the QPEs in Ansky, it remains qualitative and will require confirmation through future numerical simulations. We also caution that the presence of a pre-existing AGN disk could introduce additional complexity.

Importantly, we do not rule out alternative models. In particular, mass transfer scenarios could still be relevant if the transferred material interacts with the disk rather than accreting directly onto the SMBH, producing similar effects to those expected from stream-disk collisions. Similarly, disk instability models, especially those involving magnetically-supported radiation-pressure instabilities, remain promising alternatives, as they naturally accommodate variability in recurrence times without requiring orbital modulation. However, such models face challenges in reproducing key features of the QPE flare profiles, such as their rapid rise and slower decay.

Further monitoring is required to fully understand the nature and evolution of the QPEs in Ansky. Its high brightness makes it an ideal source for studying the mechanisms driving QPEs, and the ongoing, approved monitoring with \textit{NICER} and \textit{XMM-Newton} will provide valuable data to track the continued evolution of the period. These observations will be key to constraining physical models that can explain Ansky's peculiar behaviour.

\begin{acknowledgements}
We thank the referee for constructive feedback that helped improving the manuscript. The authors are grateful to the participants of the X-ray Quasi-Periodic Eruptions \& Repeating Nuclear Transients conference (16–19 June 2025, ESAC, Madrid) for valuable discussions.    
We acknowledge funding from ANID programs:  Millennium Science Initiative Program NCN$2023\_002$ (LHG, JCu, PA, PL), Millennium Science Initiative, AIM23-0001 (LHG, PA), FONDECYT Iniciación 11241477 (LHG), FONDECYT Regular 1251444 (JCu), 1241422, 1241005 (PA), 1230345 (CR), ANID BASAL project FB210003 and the China-Chile joint research fund (CR), and CAV, CIDI N. 21 U.de Valparaíso, Chile (PA). GM acknowledges funding from the Spanish MICIU/AEI/10.13039/501100011033 and ERDF/EU grants PID2020-115325GB-C31 and PID2023-147338NB-C21. RA was supported by NASA through the NASA Hubble Fellowship grant \#HST-HF2-51499.001-A awarded by the Space Telescope Science Institute, which is operated by the Association of Universities for Research in Astronomy, Incorporated, under NASA contract NAS5-26555. JC's research was partially supported by the Munich Institute for Astro-, Particle and BioPhysics (MIAPbP) which is funded by the Deutsche Forschungsgemeinschaft (DFG, German Research Foundation) under Germany's Excellence Strategy – EXC-2094 – 390783311. MG is funded by Spanish MICIU/AEI/10.13039/501100011033 and ERDF/EU grant PID2023-147338NB-C21.
\end{acknowledgements}


\bibliographystyle{aa}
\bibliography{refs} % if your bibtex file is called example.bib

\end{document}